\documentclass[preprint,authoryear,12pt]{elsarticle}
\usepackage{graphicx}
\usepackage{amssymb}
\usepackage{amsmath}
\usepackage{natbib}
\journal{New Astronomy Reviews}

\begin{document}

\begin{frontmatter}

\title{Formation of Black Hole Low-Mass X-ray Binaries}

\author{Xiang-Dong Li}
\ead{lixd@nju.edu.cn}
\address{Department of Astronomy and Key Laboratory of Modern Astronomy
and Astrophysics, Nanjing University, Nanjing 210046, China}


\begin{abstract}
The majority of known Galactic black holes reside in low-mass X-ray
binaries. They are rare and fascinating objects, providing unique
information on strong gravity, accretion disc physics, and stellar
and binary evolution. There is no doubt that our understanding of
the formation of black hole low-mass X-ray binaries has
significantly advanced in the past decade. However, some key issues
are still unresolved. In this paper we briefly summarize the
observational clues and theoretical progress on the formation of
black hole low-mass X-ray binaries.
\end{abstract}


\end{frontmatter}

% \linenumbers

\section{The Galactic black hole X-ray binaries}
\label{}

Stellar evolution predicts that black holes (BHs) form from massive
stars with mass $\gtrsim 20 M_{\odot}$ \citep[e.g.,][and references
therein]{fry12}, and that there are more than $10^8$ BHs in the
Galaxy \citep[e.g.,][]{vdh92}. However, only BHs in X-ray binaries
can be detected through accretion.

Currently there are 24 X-ray binary systems in which the BHs have
been confirmed via dynamical observations \citep{rem06,cha06,cas14}.
Five of them possess O or B type companion stars and belong to high-
and intermediate-mass X-ray binaries. In the remaining 19 binaries
the companion stars are of relatively low mass ($\lesssim 2
M_{\odot}$), and these binaries are transient X-ray sources with
episodic outbursts, which are likely to be caused by thermal and
viscous instabilities in an accretion disc around the BH
\citep{vp96,king96,dub99,las01}\footnote{While the thermal/viscous
instability can explain the main features of X-ray outbursts in a
general way, there are phenomena which are clearly not in agreement
with that picture. See \citet{las01} and \citet{m14} for more
detailed discussion.}. The main properties of the BH (candidate)
low-mass X-ray binaries (LMXBs) can be summarized as follows.

\noindent (1) The orbital periods

While a few BHLMXBs (e.g., XTE J1550$-$564, GX339$-$4, GS2023+338,
GRS1915+105, GS 1354$-$64, GROJ1655$-$40, 4U1543$-$47) have orbital
periods $P_{\rm orb}$ longer than 1 day, others are compact binaries
with $P_{\rm orb}<1$ day. MAXI J1659$-$152 is the shortest period BH
candidate binary known to date, with a period of $2.414 \pm 0.005$
hr \citep{kuu13}. Observations also show that XTE J1118+480 and
A0620$-$00 are undergoing rapid orbital decay at a rate
$\dot{P}_{\rm orb}=-1.90\pm 0.57~\rm ms\,yr^{-1}$ and $-0.60\pm
0.08~\rm ms\,yr^{-1}$, respectively \citep{gonz12,gonz13}.

\noindent (2) The companion stars

In the majority of the binaries, the donor companions are dwarf
stars with spectral types ranging from A2V to M1V; there are a few
cases in which the donors are giants or subgiants (e.g., K1/5III for
GRS1915$+$105, K0-K3IV for V404 Cyg, B9III for XTE J1819.3$-$2525,
F3-G0V for GRO J1655$-$40, and K2/4IV for XTE J1550-564)
\citep[][and references therein]{cas14}. These properties suggest
that most of them are low-mass stars (less than $1\, M_{\odot}$).
For those in which the lithium abundance was measured (GS 2000$+$25,
A0620$-$00, and Nova Mus 91), observations show the excess above the
solar value by a factor of about $20-200$ \citep{mar92,mar94,mar96}.

\noindent (3) The X-ray luminosities

A typical BHLMXB spends most of its time in quiescence,  and
occasionally exhibit outbursts during which its luminosities
increase by several orders of magnitude. During outbursts the X-ray
luminosities rise to $\sim 10^{37} -10^{39}$ ergs$^{-1}$; between
outbursts, they remain in a ``quiescent" state, with typical X-ray
luminosities below $\sim 10^{32}$ ergs$^{-1}$, except
 V404 Cyg, which has a much brighter quiescent luminosity,
around a few times $\sim 10^{33}$ ergs$^{-1}$
\citep{chen97,gar01,wu10}.

\noindent (4) The BH masses and spins

Astrophysical BHs are characterized by  mass and spin. Dynamical
mass determinations have been made in 17 BHLMXBs, with the BH mass
ranging from $\sim 2.7\, M_{\odot}$ to $\gtrsim 15\, M_{\odot}$
\citep[][and references therein]{cas14}. Statistical studies on the
mass distribution of neutron stars (NSs) and BHs
\citep{bai98,ozel10,farr11} suggest the presence of a mass gap or
dearth of compact objects in the interval $\sim 2-5 M_{\odot}$
\citep[see however,][]{kre12}. Numerical simulations of supernova
(SN) explosions typically generate a continuous distribution of BH
masses that decays with mass \citep{fry99,fry01}. The paucity of BHs
with masses less than $5 M_{\odot}$ is probably related to the
physics of SN explosions that lead to the formation of BHs
\citep{fry12,bel12}.

The spins of  nearly 20 BHs have  been estimated \citep[][and
references therein]{mcc14,fab14}, by means of X-ray
continuum-fitting \citep{zha97,dav05} and  modeling relativistic
reflection \citep{fab89}. The measured spins for BHs in LMXBs range
from $0.12\pm 0.19$  \citep[for A0620$-$00,][]{gou10} to $>0.98$
\citep[for GRS 1915+105,][]{mcc06,mil13}. However, one needs to
caution that both methods are model dependent to some extent, and
the results obtained for the same sources do not always agree each
other.

\section{The standard model for the formation of BHLMXBs}

All Galactic BHLMXBs dynamically confirmed are found in the Galactic
field with no apparent nearby dense clusters, so dynamical
interactions (i.e., tidal capture and exchange encounter) are not
likely to occur during their formation processes\footnote{There are
several quiescent BH candidates discovered from radio studies in
globular clusters \citep{str12,cho13}, but these are not yet
dynamically confirmed.}. The standard scenario for the LMXB
formation \citep{vdh83,bha91,tau06} starts with a detached
primordial binary with an extreme mass ratio in a relatively wide
orbit. Since most of the BHLMXBs are compact binaries with orbital
periods less than 1 day, there must exists a process that leads to
significant orbital decay of the primordial binary. After the
massive primary star evolves off the main sequence, its radius
rapidly increases, initiating Roche lobe overflow. The mass transfer
proceeds on a dynamical timescale since the primary star is much
more massive than the secondary. The secondary star spirals into the
stellar envelope when the massive star engulfs it. This is the
so-called common envelope (CE) phase \citep{pac76}. The binary orbit
changes dramatically since the orbital energy is lost due to
friction between the secondary star and the envelope of the primary,
and part of the lost orbital energy is used to eject the envelope.
Depending on whether the energy dissipated is enough to expel the
envelope, the CE evolution will result in either a merger of the
secondary star and the core of the primary, or a close binary
consisting of the secondary star orbiting around the naked core of
the primary star \citep[see][for recent reviews of CE
evolution]{taam10,iva13}. In the latter case, as the duration of the
CE phase is very short (probably less than a few hundred years), the
secondary, which is still relatively unevolved, is assumed to remain
intact. The core continues its evolution towards core collapse to
form a BH. The binary orbit is altered due to mass loss during the
core collapse and possibly an natal kick is imparted to the BH. If
the binary survives all these stages,  the secondary star will
evolve and overflow its Roche lobe, due to either nuclear expansion
or orbital angular momentum loss caused by magnetic braking (MB)
and/or gravitational radiation. Mass transfer onto the BH leads to
the formation of a LMXB. In the former case the binary will evolve
with increasing orbital periods, while in the latter, a converging
system is formed until the donor becomes degenerate. If the mass
transfer proceeds at a sufficiently low rate, thermal and viscous
instability may occur in the accretion disc around the BH, leading
to transient behavior. Based on the spatial distribution and
recurrence times of known BHLMXBs, \citet{wij96} and \citet{rom98}
estimated that there may be $\sim 2000$ such systems in the Galaxy.

There are several unresolved issues in the standard scenario, which
are outlined as follows. (1) The secondary star, because of its low
mass, may not have enough energy to eject the envelope of the BH
progenitor during the CE phase, unless a significant fraction of the
envelope has been previously lost through a very efficient stellar
wind; (2) The primary star may lose a large fraction of its mass
during the SN explosion that produced the BH, and that may disrupt
the binary. To show the former in some detail, we use the standard
energy conservation equation \citep{web84} to deal with the CE
evolution,
\begin{equation}
\alpha_{\rm CE}(\frac{GM_{\rm 1,f}M_{2}}{2a_{\rm f}}
-\frac{GM_{\rm 1,i}M_{2}}{2a_{\rm i}}) =-E_{\rm bind},
\end{equation}
and
\begin{equation}
E_{\rm bind} = -\frac{GM_{\rm 1,i}M_{\rm 1,env}}{\lambda R_{\rm 1,lobe}},
\end{equation}
where $M_1$ and $M_2$ are the primary and secondary masses
respectively, $a$ the orbital separation of the binary, $ M_{\rm
1,env} $ the mass of the primary's envelope that is ejected from the
system during the CE evolution, $ R_{\rm 1,lobe} $ the Roche-lobe
radius of the primary at the onset of overflow, and the indices i
and f refer to the initial and final stages of the CE evolution
respectively. The parameter $ \lambda $ includes the effect of the
mass distribution within the envelope, and the (possible)
contribution from the internal energy \citep{kool90,dewi00}, and $
\alpha_{\rm CE} $ is the CE efficiency with which the orbital energy
is used to unbind the stellar envelope. As pointed out by
\citet{jus06}, Eqs. (1) and (2) lead to the lower limit for the
parameter $\lambda$ in order to avoid a merger of the primary's core
and the secondary
\begin{equation}
\lambda\gtrsim 0.15\alpha^{-1}_{\rm CE}(\frac{R_{\rm 1, max}}{2300R_{\odot}})^{-1}
(\frac{M_{\odot}}{M_2})^{1/2},
\end{equation}
where $R_{\rm 1, max}$ is the maximum stellar radius of the BH
progenitor. Thus, for a secondary with $M_2 \lesssim 1\,M_{\odot}$,
$\lambda$ must be at least $\sim 0.15$. However, calculations by
\citet{pod03} show that stars of mass $M_1>25\,M_{\odot}$ have
$\lambda < 0.1$ for all $R_1 > 300R_{\odot}$, even when all the
energy of recombination for the ionized species in the envelope that
can be liberated to aid the CE ejection process is included.
Therefore, ejection of the massive envelope of a BH progenitor by a
low-mass star proves rather difficult if not impossible. This fact
has been noted by \citet{por97}. Assuming $\alpha_{\rm CE}\lambda=
0.5$, these authors showed that the parameter space of the
semi-major axis of the primordial binary for the survival of the
spiral-in process is extremely small, so that the predicted birth
rates are about two orders of magnitude lower than derived from
observations. \citet{kal99} also argued that abnormally high values
of the efficiency parameter $\alpha_{\rm CE}$ are required to gain
any agreement with the observationally inferred BHLMXB birth rate
\citep[see also][for similar investigations]{kiel06,yun08}.

\section{Modified models}

\subsection{Intermediate-mass companions}

As seen from Eq.~(3), if the secondary stars are of intermediate
mass (i.e. $\sim 3-5\,M_{\odot}$), they would be more likely to
eject the envelope. This mens that at lease part of LMXBs may evolve
form intermediate-mass X-ray binaries (IMXBs), when the donor mass
becomes $<1\,M_{\odot}$ due to mass transfer. During the IMXB phase,
since mass is transferred from the less massive secondary to the
more massive BH (usually of mass $>5\,M_{\odot}$), the binary orbit
increases accordingly. This can naturally explain the formation of
the long-period systems like GRS 1915+105 \citep{pod03}, but poses a
problem for the short-period systems. Obviously an efficient
mechanism for orbital angular momentum loss is required for the
formation of compact BHLMXBs.

Although intermediate-mass stars are not thought to be subject to
MB, \citet{jus06}  proposed that the secondary stars may be
initially Ap or Bp stars, which are known to possess anomalously
strong ($\sim 10^2-10^4$ G) magnetic fields \citep{moss89,bra04}.
Combining with the assumption that a substantial stellar wind may be
driven from the donor star  by the flux of X-ray radiation that is
produced by accretion onto the BH, they showed that
intermediate-mass donor stars with an anomalous magnetic field (with
$B\gtrsim 10^3$ G) can be braked sufficiently to form short-period
systems.

Chen \& Li (2006) proposed an alternative way of angular momentum
loss from IMXBs. They assumed that  a small fraction ($\delta$) of
the transferred mass from the donor star during the RLOF may not
accrete onto the BH, but leave the binary and form a circumbinary
disc. The tidal torques exerted by the disc can effectively drain
orbital angular momentum from the binary \citep{spr01,taam01}. It
was shown that  with $\delta\sim 0.01-0.1$ (depending on the initial
orbital periods), a circumbinary disc can cause secular orbital
shrinking, leading to the formation of compact BHLMXBs; for smaller
$\delta$, the orbits always expand during the evolution. It is
interesting to note that \citet{muno06} have detected excess
mid-infrared radiation from A 0620$-$00 and XTE J1118+480 with an
emitting areas $\sim 2$ times larger than the binary orbital
separations, providing possible evidence of the existence of a
circumbinary disc around some BHLMXBs. However, \citet{gal07}
suggested that synchrotron emission from a partially self-absorbed
outflow might be responsible for the observed mid-IR excess, in
place of, or in addition to, thermal emission from the circumbinary
material.

Support of the IMXB models comes from the CNO-processed material
seen at the surface of XTE J1118+480 from the ultraviolet spectra
\citep{has02}. This means that the companion star must be partially
nuclear-evolved and have lost its outer layers, exposing inner
layers which have been mixed with the CNO-processed material from
the central nuclear-burning region. Mass transfer must therefore
have been initiated from a somewhat evolved donor of initial mass
$M\gtrsim 1.5\,M_{\odot}$ \citep{fra09}.

However, the effective temperatures of the model donor stars are
significantly higher than for those of the observed donor stars in
BHLMXBs \citep{jus06}. The problem results from the fact that, even
though the initially intermediate-mass star loses much of its mass
fairly rapidly, the star is still somewhat evolved chemically (both
in helium and in CNO abundance) by the time its mass has been
reduced to $\sim 1\,M_{\odot}$. This keeps the donor stars hotter
than observed.

Another important clue on the initial mass of the secondary can be
derived from the spin of the BH. As the BH accretes mass and angular
momentum, its spin parameter ($a\equiv cJ/GM^2$) increases according
to \citep{bar70,tho74,king99},
\begin{equation}
a=\left(\frac{2}{3}\right)^{1/2}\left(\frac{M_{\rm BH}^0}{M_{\rm BH}}\right)
\left\{
4-\left[18\left(\frac{M_{\rm BH}^0}{M_{\rm BH}}\right)^2-2\right]^{1/2}
\right\}
\end{equation}
for $M_{\rm BH}<\sqrt{6}M_{\rm BH}^0$. Here $M_{\rm BH}^0$ and
$M_{\rm BH}$ are the initial and current masses of the BH,
respectively. If the BH spin in LMXBs is acquired through accretion
after its formation, one may estimate the accreted mass ($\Delta
M=M_{\rm BH}^0-M_{\rm BH}$) by the BHs and the possible initial mass
range the secondary from the measured BH mass and spin. That can
provide useful constraints on the properties of the progenitor
binaries. This idea can be tested for any arbitrary BHLMXB by
comparing the derived BH spin based on the orbital period and the
spectral class of the donor star with the measured one. It may also
help recover the ``true" birth mass distribution for BHs from the
current one \citep[e.g.,][]{fra15}.
%The measured values of $(M_{\rm BH},a)$ for
%A0620$-$00 and XTE J1550$-$564 are $(6.6\pm 0.25\,M_{\odot},0.12\pm
%0.19)$ \citep{can10,gou10} and  $(9.1\pm
%0.6\,M_{\odot},0.34^{+0.2}_{-0.28})$ \citep{oro11,ste11},
%respectively. One can derive $\Delta M< 0.6 M_{\odot}$ for
%A0620$-$00, and $\Delta M< 1.6 M_{\odot}$ for XTE J1550$-$564 from
%Eq.~(4). Combining with the secondary masses $0.68\pm
%0.18\,M_{\odot}$ \citep{gel01} and $0.3\pm 0.07\,M_{\odot}$
%\citep{oro11}, these values imply that the masses of the secondary
%stars of A0620$-$00 and XTE J1550$-$564 at the beginning of mass
%transfer should be less than $\sim 1 \,M_{\odot}$ and $\sim 1.9
%\,M_{\odot}$, respectively

Therefore it seems that, the survival of CE evolution seems to
require that the secondary be initially of intermediate mass, while
the apparent donor spectral classes imply that the secondary would
be of low mass at the onset of the Roche lobe overflow. A plausible
solution to this puzzle is that the secondary may be initially an
intermediate-mass star, but lose a significant part of its mass
after the formation of the BH, probably due to the ejecta impact
during an aspherical SN \citep{li08}\footnote{ The supersolar
abundances of Mg, Al, Ca, Fe and Ni in the atmosphere of the
companion stars of GRO J1655$-$40 \citep{isr99,gonz08} and XTE
J1118+480 \citep{gonz06} suggest that these BHs are likely to form
in a SN event. }. The ablated secondary star could then become a
low-mass star, and MB would shrink the binary orbit and drive mass
transfer, producing a compact BHLMXB.

\subsection{CE evolution}

\citet{pod03} found that the values of $\lambda$ that are
appropriate to stars in the mass range $(25-45) M_{\odot}$ and the
later phase of their evolution (i.e., in or beyond the Hertzprung
gap) lie in the range of $0.01\lesssim\lambda\lesssim 0.06$. These
values were obtained under the following assumptions: (1) the
definition of $\lambda$ includes the gravitational binding energy,
the thermal and the ionization energy within the envelope
\citep{han94,dewi00} ; (2) the core mass in the primary star is
defined as the central mass that contains $1 M_{\odot}$ of hydrogen,
and (3) the primary star loses mass through stellar wind at a rate
according to the prescription of \citet{nie90}. Since CE evolution
plays a key role in the formation of these systems, in the following
we will discuss each point in some detail.

Besides the thermal and ionization energy, other energy sources may
also contribute to the envelope ejection. One of them is nuclear
fusion \citep{iva02}. This can occur during the slow merger of a
massive primary with a secondary of mass $\sim 1-3 M_{\odot}$, when
the primary has already completed helium core burning. \citet{pod10}
suggested that, once the inspiraling secondary fills its own Roche
lobe during the CE phase, a stream of hydrogen-rich material from
the secondary can penetrate deep into the primary's core, mixing
hydrogen into the helium-burning shell, and leading to a
thermonuclear runaway. The released nuclear energy ($\gtrsim$ a few
times $10^{50}$ ergs) during this explosive hydrogen burning could
exceed the binding energy of the helium shell, and result in the
explosive ejection of both the hydrogen and the helium layers,
producing a close binary containing a CO star and a low-mass
companion. The subsequent evolution then leads to a compact BHLMXB.
This scenario also suggests that the formation of BHs in this kind
of systems could have been accompanied by long-duration gamma-ray
bursts.

\citet{iva11} pointed out that the usually adopted energy criterion for the
CE phase is not sufficient for an envelope ejection.
They argued that mass outflows are likely to occur during the slow spiral-in
stage, and the condition for such outflows is to balance
the orbital energy with the enthalpy of the envelope rather than merely the internal energy.
This is to add a pressure/density ($P/\rho$) term in the binding energy equation
\begin{equation}
E_{\rm bind}=-\int_{\rm core}^{\rm
surface}[\Psi(m)+\epsilon(m)+P(m)/\rho(m)]{\rm d}m,
\end{equation}
where $\Psi$ is the gravitational potential energy, and $\epsilon$
is the specific internal energy. Since $P/\rho$ is non-negative, the
condition to start outflows occurs before the envelope's total
energy becomes positive. The value of $\lambda$ is then larger than
usually estimated by a factor of $\sim 2 - 5$, which might help
allow for a low-mass ($\lesssim 1\,M_{\odot}$) companion to survive
from the CE phase, even if the primary star is initially as massive
as $30\,M_{\odot}$. There are arguments both for and against this
``enthalpy" formalism from theoretical points of view, as outlined
by \citet{iva13}. Different formalisms of $\lambda$ may be
distinguished by applying them to the formation of specific binaries
and comparing the results with observations\footnote{When
constructing the evolutionary history of IC 10 X$-$1, one of the
three observed BH X-ray binaries that are known to host a Wolf-Rayet
star as the mass donor \citep{cla04}, \citet{wong14} obtained
constraints on the physics of the CE event, and found that the
``enthalpy" formalism is the only one that can explain the existence
of IC 10 X$-$1 without the need of invoking unreasonably high CE
efficiencies $\alpha_{\rm CE}$. Unfortunately, the adopted $\sim 30
M_{\odot}$ BH mass is almost certainly incorrect, because it was
derived from the radial velocity curve from the wind emission lines,
under the unrealistic assumption that the wind is spherically
symmetric \citep[see][for a discussion on this
subject]{vk93,mll14}.}. Obviously a proper estimate of the energy
requirement for the CE ejection is critical for the formation of not
only BHLMXBs but also all kinds of post-CE binaries.

Another important physical question in calculating $E_{\rm bind}$ is
how to determine the boundary between the remaining ``core" and the
ejected ``envelope" of the star. There are different definitions of
the boundary in the literature, which are (1) related to the nuclear
energy generation within the hydrogen burning shell, (2) based on
the the hydrogen abundance, and (3) connected with thermodynamic
quantities \citep{iva13}. \citet{tau01} have shown that, depending
on the definition of the core-envelope boundary  and the treatment
of the role of internal thermodynamic energy, the binding energy
parameter $\lambda$ can vary by two orders of magnitude (from 0.02
to 3.50) for the a $20\,M_{\odot}$ star at the tip of red giant
branch. This sensitivity results from the fact that the binding
energy within the hydrogen burning shell greatly exceeds that of the
outer convective envelope.

\citet{ivan11} suggested that
every giant has a well-defined post-CE remnant after it has been
thermally readjusted. The mass $m_{\rm cp}$ of the remnant is most likely given by the
divergence point,
which is best approximated by the point in the
hydrogen burning shell that had maximal compression (local
sonic velocity)  prior to CE.
If the remnant mass remains above or below $m_{\rm cp}$, then the star expands
to overflow its Roche lobe or contracts during its thermal readjustment.
Thus $m_{\rm cp}$ could be regraded as the bifurcation point which defines where the spiral-in
stops.

\begin{figure*}[htbp]
\centering
%\plottwo{f1a.eps}{f1b.eps}
\includegraphics[height=0.35\textwidth]{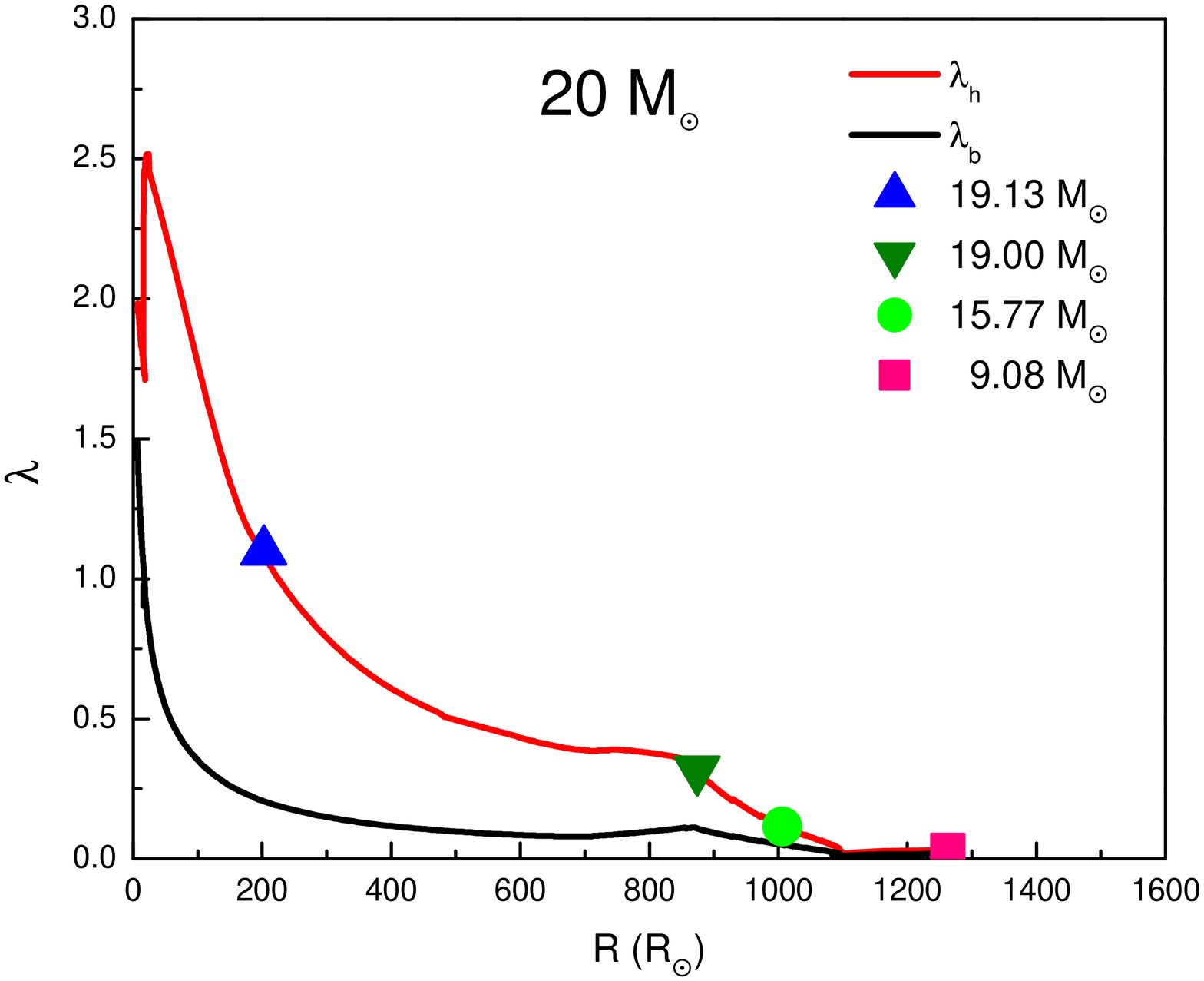}
\includegraphics[height=0.35\textwidth]{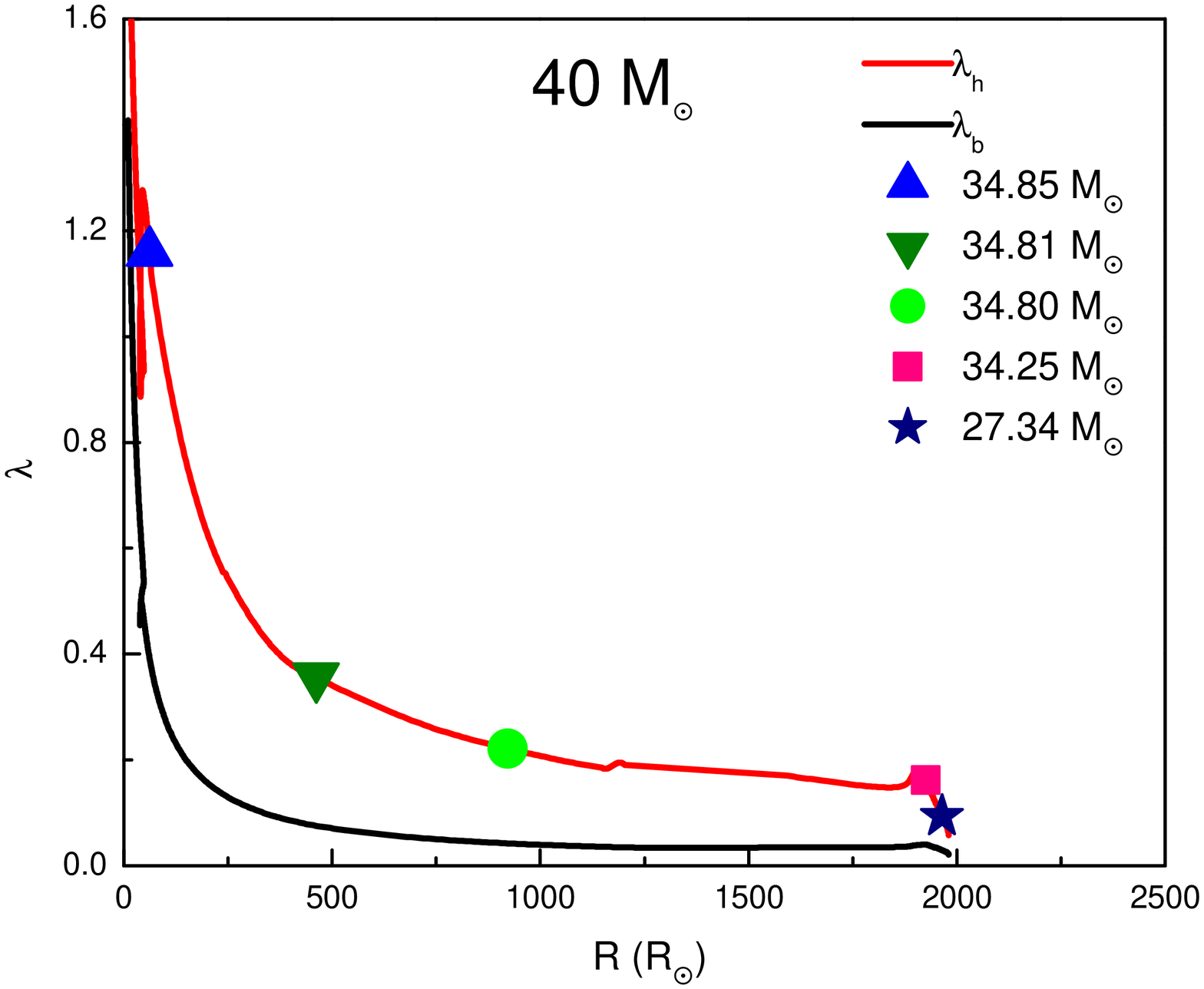}
\caption{Dependence of the parameter $\lambda$ on the stellar radius $R$
for stars with initial
mass of $20\,M_{\odot}$ (left) and $40\,M_{\odot}$ (right).
The red and black lines denote $\lambda$ with and without including
the contribution of the $P/\rho$ term ($\lambda_{\rm h}$ and
$\lambda_{\rm b}$, respectively) in the binding energy. Also shown are the stellar mass
at different stages of the evolution.
\label{}}
\end{figure*}

If the orbit of the primordial binary is initially so wide that CE
is initiated very late in the evolution of the primary star, the
star may have lost most of its envelope in stellar winds, which
means that there is not much orbital energy required for successful
envelope ejection even with a low-mass companion star. \citet{wik14}
therefore argued that CE evolution is not crucial in the formation
of Galactic BHLMXBs, and stars with mass $\sim 1\,M_{\odot}$ are the
most likely companions in BHLMXBs. It should be noted that this
scenario may work for very massive primary stars, since the stellar
wind prescriptions should not significantly affect the stellar
structure and the compact object remnant masses for stars with
initial mass $\lesssim 30\,M_{\odot}$. It also requires that the
progenitor binaries should have very wide orbits, so the parameter
space for the formation of BHLMXBs could be quite small.

In Fig.~1 we compare the values of $\lambda$ with and without
including the contribution of the $P/\rho$ term in the binding
energy (denoted as $\lambda_{\rm h}$ and $\lambda_{\rm b}$,
respectively) as a function of the stellar radius (Wang \& Li 2015,
in preparation), adopting the core-envelope boundary suggested by
\citet{ivan11}. The left and right panels are for $20\,M_{\odot}$
and $40\,M_{\odot}$ stars, respectively. We also consider mass loss
due to stellar winds as in \citet{xu10} (which is actually same as
in Hurley et al. 2000), and denote the stellar mass during its
evolution. It is seen that,  in both cases, $\lambda_{\rm h}\gtrsim
0.15$ for a rather wide range of stellar radius (except near the end
of the evolution), while $\lambda_{\rm b}< 0.15$ except for small
radius.

\subsection{Other models}
\citet{egg86} suggested that the progenitor of a BHLMXB is a triple
star, in which a massive close binary is accompanied at a large
distance by a low-mass star. After the evolution of the close binary
into an ordinary X-ray binary, the compact object is engulfed by its
expanding massive companion, and spirals in to settle at its center.
The resulting Thorne-\.Zytkow object \citep[T\.ZO,][]{tho75,tho77}
gradually expands until it attains the size of the low-mass star's
orbit. Then a second spiral-in phase ensues, leading to the
formation of a low-mass close binary.

\citet{pod95} proposed that during the evolution of T\.ZOs the
central NS may be converted into a BH by intense accretion, and part
of the envelope may collapse into a massive disc, which may become
gravitationally unstable and lead to formation of low-mass stars or
planets. Further evolution would be similar as standard LMXBs.

\citet{iva06} argued that some BHLMXBs could harbour pre-main
sequence donors. For a binary with an extreme mass ratio, the
primary star evolves rapidly to explode and create a BH, while its
low-mass companion has not reached the zero-age main sequence at
that time. As a pre-main sequence star usually possesses strong
magnetic fields, during its contraction toward the main sequence, MB
dissipates the orbital angular momentum, and drives mass transfer.
Because pre-main sequence stars are cooler than main sequence stars,
the temperature issue in the IMXB models would be avoided.

The aforementioned models can also explain the enhanced lithium
abundance measured in several BHLMXBs (though in different ways),
because either T\.ZOs provide an ideal environment for the
production of $^7$Li by the $^7$Be-transport mechanism, or pre-main
sequence stars can have high (up to primordial) Li abundance since
no nuclear burning occurs in them. The predicted BHLMXBs should be
young systems or subject to small kicks during the formation of BHs,
and therefore have relatively low space velocities and a very small
Galactic scaleheight. However, none of the Galactic BH transients
are associated with or found nearby star-forming regions. It is also
difficult for these models to explain the spatial distribution BH
binaries like GROJ1655$-$40 \citep{wil05}, XTE J1118+480
\citep{fra09} and MAXI J1659$-$152 \citep{yam12}, which are thought
to have received rather large kick velocities ($\sim 100-200$
kms$^{-1}$) at birth. They might be suitable for V404 Cyg, which has
a peculiar velocity of about 40 kms$^{-1}$ \citep{mil09}, implying
that its BH did not receive a velocity kick larger than this value.

\section{Discussion and conclusions}

BHLMXBs are exotic objects because they are very difficult to form;
they are also fascinating objects because they offer the best
opportunity to study not only astrophysical phenomena associated
with extreme gravity, but also binary evolution and SN mechanisms.
We have briefly reviewed the progress made over the past decade in
theoretical investigations on the formation of BHLMXBs, in
particular short-period systems. Understanding these binaries is
difficult, since some fundamental issues have not been resolved,
including

\noindent (1) What is the mass range of the BH progenitor stars?

\noindent (2) Do the mass gap in BHs really exist? If it does, what's the physics behind it?

\noindent (3) What is the distribution of the kick imparted on
newborn BHs?

\noindent (4) How to properly estimate the total binding energy within the
stellar envelope during CE evolution? Do we really need an initially
intermediate-mass companion star for BHLMXBs?

\noindent (5) What's the dominant mechanism of orbital angular
momentum loss in BHLMXBs? How can it account for the rapid orbital
decay found in  XTE J1118+480 and A0620$-$00?

In a recent work, \citet{koc14} suggested that stars of mass $16.5-
25\,M_{\odot}$, which have not been observed as the progenitors of
Type IIP SNe, die in failed SNe. Such failed SNe eject their
hydrogen envelopes in a weak transient, leaving a BH with the mass
of the star's helium core $(5-8)\,M_{\odot}$. This may explain the
typical masses of observed BHs and the gap between NS and BH masses
without any fine-tuning of stellar mass loss, or the SN mechanism.
Compared with more massive stars, these stars have relatively higher
$\lambda$ (see Fig.~1), which may also help eject the envelope for a
low-mass companion star.

Obviously thorough population synthesis calculations are required to
investigate the evolutionary history of BHLMXBs by incorporating
various factors aforementioned in a more self-consistent way.

\section*{Acknowledgements}
This work was supported by the NSFC (under grant numbers 11133001 and 11333004)
and the Strategic Priority Research Program of CAS (under grant number XDB09000000).

\end{document}